# Distributed Power Control Schemes for In-Band Full-Duplex Energy Harvesting Wireless Networks

Rojin Aslani and Mehdi Rasti

*Abstract*—This paper studies two power control problems in energy harvesting wireless networks where one hybrid base station (HBS) and all user equipments (UEs) are operating in in-band full-duplex mode. We consider minimizing the aggregate power subject to the quality of service requirement constraint, and maximizing the aggregate throughput. We address these two problems by proposing two distributed power control schemes for controlling the uplink transmit power by the UEs and the downlink energy harvesting signal power by the HBS. In our proposed schemes, the HBS updates the downlink transmit power level of the energy-harvesting signal so that each UE is enabled to harvest its required energy for powering the operating circuit and transmitting its uplink information signal with the power level determined by the proposed schemes. We show that our proposed power control schemes converge to their corresponding unique fixed points starting from any arbitrary initial transmit power. We will show that our proposed schemes well address the stated problems, which is also demonstrated by our extensive simulation results.

*Index Terms*—Cellular networks; resource allocation; distributed power control; energy harvesting; in-band full-duplex.

## I. Introduction

Today, tremendous growth of network traffic increases energy consumption and carbon emissions in wireless networks, which is a major challenge against the goal of green communications [1]. Energy efficiency is a key issue in today's networks due to the limitation of energy resources. Several studies have been conducted on energy efficiency in wireless networks, including designing energy efficient MAC protocol [2], transmission scheduling [3] and channel sensing [4]. While all of the above techniques optimize and adapt the energy usage to maximize the lifetime of battery-powered nodes, the lifetime remains bounded and finite [5].

Energy harvesting which refers to the capability of converting the harvested energy from energy sources (light, vibration, body heat, foot strike, radio frequency signals) into electricity [6], is a new technique recently employed to address the problem of node's finite lifetime. The energy harvesting capability allows the wireless nodes to harvest the energy required for their information transmission. Consequently, energy harvesting wireless networks (EHWNs) have immediately found their applications in various forms, such as device to device (D2D) communications [7], wireless sensor networks [8] and cellular networks [9]. One of the important issues that arises in 5G wireless communications is to improve the energy efficiency by using the energy harvesting techniques [10].

R. Aslani and M. Rasti are with the Dept. of Computer Engineering and Information Technology, Amirkabir University of Technology, Tehran, Iran (email: {rojinaslani, rasti}@aut.ac.ir).

Architecture of an EHWN typically consists of three major components: the base station, the energy sources, and the user equipments (UEs). The base station is responsible for communication among the UEs. The energy sources can be either ambient sources such as wind, light and ambient signal interference, or dedicated energy transmitters. To control the harvested energy, the UEs may adopt either the harvest-use method or the harvest-store-use method. In the harvest-use method, the harvested energy is immediately used to power the UE. In the harvest-store-use method, the UE is equipped with an energy storage and whenever the harvested energy is more than that of current UE's consumption, the excess energy is stored for future use [6].

There have been recent research efforts on designing resource allocation schemes for wireless networks with energy harvesting capable UEs [11]-[14]. In [11], a joint downlink energy transfer and uplink information transmission problem is formulated to maximize the system throughput, and a low-complexity scheduling algorithm is proposed for it. In [12] and [13], the multi-user time allocation problem in a time division multiple access EHWN is considered. In [12], the authors characterize two optimization problems: total throughput maximization and total energy harvesting time minimization. The problem of maximizing the weighted sum-throughput is addressed in [13], by jointly optimizing the transmit power of access point and UEs and the time allocated to the access point and the UEs for transmitting the energy and data, respectively.

Existing related resource allocation schemes for EHWNs are mostly centralized and focus on the objective of throughput maximization. Although there has been a rich literature on designing distributed power control scheme in traditional wireless network without energy harvesting [15], [16], there exist very few distributed power control schemes for EHWNs. In [14] a distributed power control algorithm is proposed to minimize the transmit power of UEs in EHWN, where the source of energy harvesting is the ambient energy such as solar and wind. So, in [14], the amount of energy harvested by the UEs via an ambient energy source is unpredictable and it may make the harvested energy less than the required power at some UEs for transmitting their data. In fact, this gives rise to the problem of energy-infeasibility in the system, where some UEs (called energy-non-supported UEs) do not have the sufficient energy to transmit their data.

On the other hand, recently, there has been a growing interest in in-band full-duplex (IBFD) wireless networks, where the wireless node transmits and receives simultaneously in the same frequency band. However, due to the simultaneous transmission and reception at the same node, IBFD systems suffer from the self interference (SI) which is part of the transmitted signal of a full-duplex node received by itself, thus interfering with the desired signal received at the same time. Self-interference cancellation (SIC) is a key challenge for implementing IBFD communications. Various SIC techniques have been proposed in the literature (e.g., see [26] and [27]). First two pioneer practical implementations of IBFD communications have been proposed in [24] and [25]. The authors of [12] and [13] apply the IBFD technique to an EHWN, where a dual-function hybrid base station (HBS) is employed for transferring energy to all UEs, and concurrently acting as a base station to receive the UEs' information. Most of the existing related works assume that the UEs operate in half-duplex mode with a TDMA structure to transmit their information to the HBS.

In this paper, we focus on designing distributed power control schemes for in-band full-duplex energy harvesting wireless networks (IBFD-EHWNs). We assume that the energy source is dedicated, where an HBS is employed for transferring the energy signals to the UEs and concurrently receiving their information signals. We further assume that in addition to the HBS operating in the IBFD mode as in [12] and [13], the UEs are also operating in the IBFD mode (in contrast to [12] and [13] where the UEs operate in half-duplex mode). The UEs harvest energy from the downlink energy signal transmitted by the HBS and use it to power their operating circuits and transmit their information signals to the HBS in the uplink, while both the uplink and downlink transmissions use the same frequency band. We refer to such an EHWN model with dual-function HBS and UEs as IBFD-EHWN. To the best of our knowledge, no other work has focused on designing distributed power control schemes for IBFD-EHWNs. Our contributions in this paper are summarized as follows.

- We present a new system model for IBFD-EHWNs under which two power control optimization problems are formally stated; namely, (i) aggregate power minimization problem, i.e., minimizing the aggregate consumed power of the UEs (transmit and circuit power) and the power of the HBS (energy harvesting transmit signal and circuit power), subject to the energy harvesting constraint and quality of service requirement of UEs, and (ii) throughput maximization problem, i.e., maximizing the total throughput of the system subject to the energy harvesting constraint of the UEs. To the best of our knowledge, these two problems have not been considered for IBFD-EHWNs in the literature.
- For the aggregate power minimization problem, we first obtain the corresponding optimal power allocation for the UEs and the HBS in IBFD-EHWN, and then we propose an iterative power control scheme to address the problem in a distributed manner. We show that our proposed distributed power control scheme converges to its unique fixed point which corresponds to the optimal solution of the aggregate power minimization problem. By harvesting the energy from the HBS and proper adjustment of the UEs' transmit power in our proposed scheme, all UEs achieve their target signal-to-interference-plus-noise ratio (SINR) while the minimum aggregate power is consumed by UEs and HBS. The efficiency of our proposed method is demonstrated by the simulation results.
- To address the throughput maximization problem, we also propose a distributed iterative power control scheme for allocating the transmit power of the UEs and the HBS, which converges to its unique fixed point. The simulation results demonstrate that our proposed scheme enables each UE to harvest enough energy for powering its operating circuit and transmitting information signals. Moreover, our proposed method significantly improves the aggregate throughput of the UEs in IBFD-EHWNs (to the same extent of improvement achieved by the opportunistic power control (OPC) algorithm [16] in traditional cellular networks without energy harvesting capability as compared to the other power control algorithms such as the algorithm in [15]).

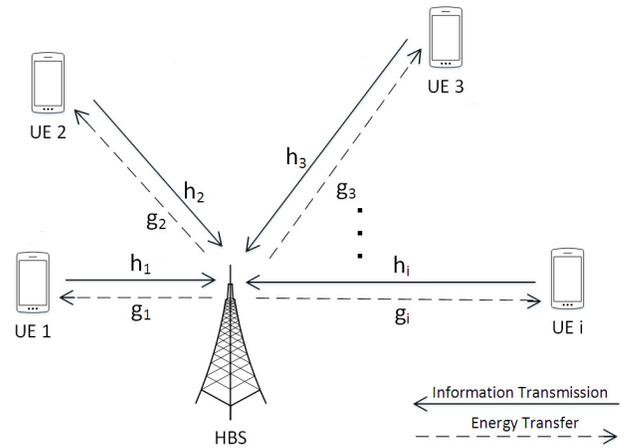

Fig. 1. EHWN system model.

The rest of this paper is organized as follows. In Section II, we introduce the system model and formally state the problems. Our distributed power control schemes for addressing the stated problems are presented and analyzed in Section III. Sections IV and V contain the numerical results and the conclusions, respectively.

## II. SYSTEM MODEL AND PROBLEM FORMULATION

### A. System Model

We consider an EHWN with $K$ UEs, forming the set $\mathcal{K} = \{1, 2, \cdots, K\}$, and one HBS. We focus on the case of IBFD-EHWN where both the HBS and UEs operate in IBFD mode (energy transfer to the UEs in the downlink and UEs data transmission in the uplink are simultaneously performed over the same frequency band). More specifically, in contrast to [14] and [22], wherein the UEs harvest energy from an ambient source and the base station only performs data communications with the UEs, we assume that the HBS is equipped with two antennas. One is for the downlink wireless energy transfer and the other one is used to receive the uplink information transmitted by the UEs. All UEs are also assumed to be equipped with two antennas, and operate in IBFD mode by utilizing one antenna for harvesting energy from the HBS in the downlink and the other antenna for transmitting information to the HBS in the uplink, simultaneously over the same frequency band. The downlink channel power gain from the HBS to the $i$th UE is denoted by $g_i$. The uplink channel power gain from the $i$th UE to the HBS is denoted by $h_i$.

During the time interval of $\Delta t$, the HBS broadcasts wireless energy by transmitting a signal with power $p^{\text{H}}$ to all UEs in the network, where $0 \leq p^{\text{H}} \leq \overline{p}^{\text{H}}$, in which $\overline{p}^{\text{H}}$ denotes the peak transmit power of the HBS. The total energy harvested by the $i$th UE from the HBS is denoted by $E_i$ and is given by

$$E_i = \mu_i g_i p^{\text{H}} \Delta t, \tag{1}$$

where $\mu_i \in (0,1)$ is a constant denoting the energy harvesting efficiency for the $i$th UE, which depends on the storage circuit of the UE such as the storage voltage [21]. In addition to the energy transferred by the HBS, the total power consumed in the HBS consists of a constant term which is associated with the operating circuit of the HBS when it is turned on. This constant term is not negligible. In this work, we adopt the linear power model as in [23], where the total power consumption at the HBS is modeled as

$$p_{\text{HBS}}^{\text{total}} = \frac{1}{\epsilon} p^{\text{H}} + p_{\text{HBS}}^{\text{cir}}, \tag{2}$$

where $\epsilon \in [0,1]$ is the power amplifier efficiency which depends on the design and implementation of the power amplifier, $p^{\text{H}}$ is the energy harvesting power transferred by the HBS, $p_{\text{HBS}}^{\text{cir}}$ is the HBS circuit power given by $p_{\text{HBS}}^{\text{cir}} = N_{\text{HBS}} p_{\text{HBS}}^{\text{Dyn}} + p_{\text{HBS}}^{\text{Sta}}$, in which $N_{\text{HBS}}$ is the number of antennas at the HBS, $p_{\text{HBS}}^{\text{Dyn}}$ is the dynamic HBS circuit power consumption corresponding to the power radiation of all circuit blocks and proportional to the number of the HBS transmit antennas, and $p_{\text{HBS}}^{\text{Sta}}$ is the static HBS circuit power spent by the cooling system, power supply, etc. Clearly, our adopted power model given by (2) considers not only the transmit power of the HBS, but also its circuit power.

We assume that the UEs adopt the harvest-use method, i.e., they have no energy source other than the transferred energy from the HBS, nor do they have batteries to store their harvested energy, and hence all harvested energy is used to power operating circuit and information transmission in the uplink. Similar to the HBS, the total power consumption at the $i$th UE is modeled as

$$p_i^{\text{total}} = \frac{1}{\epsilon} p_i^{\text{u}} + p_i^{\text{cir}}, \tag{3}$$

where $p_i^{\text{u}}$ is the transmit power of the $i$th UE, $p_i^{\text{cir}}$ is the $i$th UE's circuit power given by $p_i^{\text{cir}} = N_i p_i^{\text{Dyn}} + p_i^{\text{Sta}}$, in which $N_i$ is the number of antennas at the $i$th UE, $p_i^{\text{Dyn}}$ is the dynamic circuit power of the $i$th UE, and $p_i^{\text{Sta}}$ is the static circuit power of the $i$th UE. The total power consumed at each UE should be less than its harvested energy, i.e.,

$$p_i^{\text{total}} \leq \frac{E_i}{\Delta t} = \mu_i g_i p^{\text{H}}, \qquad \forall i \in \mathcal{K}. \tag{4}$$

The above equation is the energy harvesting constraint which ensures that the required energy for providing the total power consumed at the $i$th UE (including the required power for enabling its circuit and transmitting signals) is sufficiently harvested. By replacing $p_i^{\text{total}}$ with $\frac{1}{\epsilon} p_i^{\text{u}} + p_i^{\text{cir}}$, the energy harvesting constraint (4) is equivalently rewritten as

$$p^{\text{H}} \geq \frac{p_i^{\text{u}}}{\epsilon \mu_i g_i} + \frac{p_i^{\text{cir}}}{\mu_i g_i}, \qquad \forall i \in \mathcal{K}. \tag{5}$$

Defining the minimum required power for enabling the $i$th UE's circuit as

$$p_i^{\min} = \frac{p_i^{\text{cir}}}{\mu_i g_i}, \tag{6}$$

we can rewrite the energy harvesting constraint (4) or (5) as

$$p^{\text{H}} \geq \frac{p_i^{\text{u}}}{\epsilon \mu_i g_i} + p_i^{\min}, \qquad \forall i \in \mathcal{K}. \tag{7}$$

Due to hardware limitations, the harvested energy by the $i$th UE during $\Delta t$ cannot be greater than $\overline{E}_i$, i.e., $E_i \leq \overline{E}_i$. From $E_i \leq \overline{E}_i$ and by noting (4) and replacing $p_i^{\text{total}}$ with $\frac{1}{\epsilon} p_i^{\text{u}} + p_i^{\text{cir}}$ from (3), we have $p_i^{\text{u}} \leq \epsilon(\frac{\overline{E}_i}{\Delta t} - p_i^{\text{cir}})$. Let us define $\overline{p}_i^{\text{u}}$ as the upper limit for the transmit power of the $i$th UE which is given by

$$\overline{p}_i^{\text{u}} = \epsilon \left( \frac{\overline{E}_i}{\Delta t} - p_i^{\text{cir}} \right). \tag{8}$$

We call $p_i^{\text{u}} \leq \overline{p}_i^{\text{u}}$ the feasibility constraint of the $i$th UE's transmit power.

Let us denote the transmit power vector $\mathbf{p}$ by $\mathbf{p} = [\mathbf{p}^{\text{u}}, p^{\text{H}}]^T$, where $\mathbf{p}^{\text{u}} = [p_1^{\text{u}}, p_2^{\text{u}}, \cdots, p_K^{\text{u}}]^T$ is the UEs' uplink transmit power vector and $p^{\text{H}}$ is the HBS's downlink energy harvesting transmit power. Considering the capability of self-interference cancellation for the HBS, similar to [28], [29] and [30], we assume that the residual self-interference is proportional to the transmission power. By defining $\delta$ as the effective self-interference coefficient, $\delta p^{\text{H}}$ denotes the residual self-interference due to the simultaneous transmission of the energy signal and receiving the information signals at the full-duplex HBS. Additive white Gaussian noise (AWGN) is assumed to be present with power $\sigma^2$ at the HBS. Given the transmit power vector $\mathbf{p}$, the uplink SINR of the $i$th UE at the HBS receiver denoted by $\gamma_i$, is obtained as

$$\gamma_i(\mathbf{p}) = \frac{h_i p_i^{\text{u}}}{\sum_{j \in \mathcal{K}, j \neq i} h_j p_j^{\text{u}} + \delta p^{\text{H}} + \sigma^2}. \tag{9}$$

A given uplink SINR vector $\boldsymbol{\gamma} = [\gamma_1, \gamma_2, \cdots, \gamma_K]^T$ is feasible if there exists a feasible transmit power vector $\mathbf{p}$ (i.e., $0 \leq p^{\text{H}} \leq \overline{p}^{\text{H}}$ and $0 \leq p_i^{\text{u}} \leq \overline{p}_i^{\text{u}}$, $\forall i \in \mathcal{K}$) corresponding to $\boldsymbol{\gamma}$. Let $\widehat{\gamma}_i$ denote the minimum acceptable (target) SINR of the $i$th UE which represents its required quality of service (QoS). Given $\widehat{\gamma}_i$, we say that the QoS requirement of the $i$th UE is satisfied by the transmit power vector $\mathbf{p}$, if $\gamma_i(\mathbf{p}) \geq \widehat{\gamma}_i$, where $\gamma_i(\mathbf{p})$ is obtained from (9). The achievable instantaneous uplink transmission rate in bit/s/Hz for the $i$th UE is given by the Shannon formula as

$$R_i(\mathbf{p}) = \log \left( 1 + \frac{h_i p_i^{\text{u}}}{\sum_{j \in \mathcal{K}, j \neq i} h_j p_j^{\text{u}} + \delta p^{\text{H}} + \sigma^2} \right). \tag{10}$$

*B. Problem Formulation*

In general, it is desirable to design a scheme for power allocation so that a given objective function $f_o(\mathbf{p})$ (e.g., aggregate consumed power, or aggregate throughput of the UEs) is optimized subject to the energy harvesting constraint of the UEs, feasibility of transmit power levels for the UEs and the HBS, and the QoS requirements of the UEs. This corresponds to the following general optimization problem:

$$\underset{\mathbf{p}}{\text{optimize}} \quad f_o(\mathbf{p}) \tag{11}$$

$$\text{subject to} \quad \text{energy harvesting constraint for UEs,} \tag{11-1}$$
$$\text{feasibility of transmit power for HBS,} \tag{11-2}$$
$$\text{feasibility of transmit power for UEs,} \tag{11-3}$$
$$\text{QoS requirement for UEs.} \tag{11-4}$$

In this paper, we focus on two examples of the above general optimization problem in an IBFD-EHWN given bellow.

*1) Aggregate Power Minimization Subject to the Energy Harvesting and QoS Constraints of the UEs:* This problem is formally stated as

**Problem 1.**

$$\underset{\mathbf{p}}{\text{minimize}} \quad \sum_{i=1}^{K} p_i^{\text{total}} + p_{\text{HBS}}^{\text{total}} \tag{12}$$

$$\text{subject to} \quad p^{\text{H}} \geq \frac{p_i^{\text{u}}}{\epsilon \mu_i g_i} + p_i^{\min}, \quad \forall i \in \mathcal{K} \tag{12-1}$$

$$p^{\text{H}} \leq \overline{p}^{\text{H}}, \tag{12-2}$$

$$p_i^{\text{u}} \leq \overline{p}_i^{\text{u}}, \quad \forall i \in \mathcal{K} \tag{12-3}$$

$$\gamma_i(\mathbf{p}) \geq \widehat{\gamma}_i, \quad \forall i \in \mathcal{K} \tag{12-4}$$

where the first constraint corresponds to the energy harvesting constraint for the UEs derived in (7) and the second and third constraints correspond to the feasibility of transmit power levels for the HBS and the UEs, respectively. The last constraint corresponds to the QoS requirement for the UEs.

*2) Throughput Maximization Subject to the Energy Harvesting Constraint of the UEs:* This problem is formally stated as

**Problem 2.**

$$\underset{\mathbf{p}}{\text{maximize}} \quad \sum_{i=1}^{K} \log\left(1+\frac{h_i p_i^{\text{u}}}{\sum_{j\in\mathcal{K},j\neq i} h_j p_j^{\text{u}} + \delta p^{\text{H}} + \sigma^2}\right) \tag{13}$$

$$\text{subject to} \quad p^{\text{H}} \geq \frac{p_i^{\text{u}}}{\epsilon \mu_i g_i} + p_i^{\min}, \quad \forall i \in \mathcal{K} \tag{13-1}$$

$$p^{\text{H}} \leq \overline{p}^{\text{H}}, \tag{13-2}$$

$$p_i^{\text{u}} \leq \overline{p}_i^{\text{u}}, \quad \forall i \in \mathcal{K} \tag{13-3}$$

where the first constraint corresponds to the energy harvesting constraint for the UEs derived in (7) and the last two constraints correspond to the feasibility of transmit power levels for the HBS and the UEs, respectively.

Disabling the energy harvesting and IBFD capabilities, and assuming availability of the energy source (such as battery) for all UEs, make Problems 1 and 2 conventional power control problems in traditional wireless cellular network without energy harvesting capability (where the UEs are equipped with a permanent battery), which is formally stated and addressed in [15], [16], [17]. In more details, if we let $p^{\text{H}} = 0$ or correspondingly $\delta = 0$, and remove the constraints (12-1), (12-2) and (13-1), (13-2) from Problems 1 and 2, respectively, and consider $\overline{p}_i^{\text{u}}$ as a given maximum transmit power of an available permanent battery, then the resulting problems are addressed in a distributed manner by the target-SINR tracking power control algorithm (TPC) proposed in [15] and the OPC algorithm proposed in [16], [17], respectively. It is evident that the TPC and OPC algorithms cannot be employed for addressing Problems 1 and 2 in the IBFD-EHWN, since they are designed for a system without energy harvesting and IBFD capabilities, assuming an available permanent battery for each UE. In the subsequent section, we propose two distributed power control schemes for controlling the uplink transmit power by the UEs and the downlink energy harvesting signal power by the HBS for addressing Problems 1 and 2, respectively, assuming no permanent battery is available at the UEs, and they harvest their required energy from the HBS.

## III. OUR PROPOSED DISTRIBUTED POWER CONTROL SCHEMES AND THEIR ANALYSIS

In this section, we first propose two distributed iterative schemes for controlling the transmit power levels of the HBS and the UEs so that Problems 1 and 2 are addressed, respectively. Then, we show that our proposed schemes converge to their unique fixed points, followed by a discussion on their optimally.

*A. Our Proposed Power Control Schemes for IBFD-EHWN*

In what follows, we first propose a distributed power update function for the HBS's downlink energy harvesting transmit power. Then, we propose two distributed uplink transmit power update functions for the UEs which along with our proposed downlink energy harvesting transmit power, address Problems 1 and 2, respectively.

*1) Our proposed downlink energy harvesting power control scheme:* First, in lemma 1, we show that the optimal downlink energy harvesting transmit power for both Problems 1 and 2 is the same and is obtained as a function of the UEs' uplink transmit power. Then, we use it to propose a distributed scheme for iteratively updating the energy harvesting transmit power by the HBS.

**Lemma 1.** Let us denote the optimal solution of either Problem 1 or 2 (if feasible) by $\mathbf{p}^* = [\mathbf{p}^{*\text{u}}, p^{*\text{H}}]^T = [p^{*\text{u}}_1, p^{*\text{u}}_2, \cdots, p^{*\text{u}}_K, p^{*\text{H}}]^T$. The optimal energy harvesting transmit power by the HBS ($p^{*\text{H}}$) in both problems is given by:

$$p^{*\text{H}} = \max_{i\in\mathcal{K}}\left(\frac{p^{*\text{u}}_i}{\epsilon\mu_i g_i} + p_i^{\min}\right). \tag{14}$$

*Proof.* We prove by contradiction. If (14) does not hold, then we have $p^{*\text{H}} < \max_{i\in\mathcal{K}}\left(\frac{p^{*\text{u}}_i}{\epsilon\mu_i g_i} + p_i^{\min}\right)$ or $p^{*\text{H}} > \max_{i\in\mathcal{K}}\left(\frac{p^{*\text{u}}_i}{\epsilon\mu_i g_i} + p_i^{\min}\right)$. If $p^{*\text{H}} < \max_{i\in\mathcal{K}}\left(\frac{p^{*\text{u}}_i}{\epsilon\mu_i g_i} + p_i^{\min}\right)$, then the energy harvesting constraint (12-1) in problem 1 and (13-1) in problem 2 are not satisfied for some UEs and the system is energy-infeasible. If $p^{*\text{H}} > \max_{i \in \mathcal{K}} \left( \frac{p_i^{*\text{u}}}{\epsilon \mu_i g_i} + p_i^{\min} \right)$, then the solution is not optimal. Because, increasing $p^{\text{H}}$ will increase the aggregate power in Problem 1 and will decrease the throughput in Problem 2. This completes the proof. □

Lemma 1 enables us to propose a distributed power control scheme for controlling the HBS transmit power. Let $\mathbf{p}(t) = [\mathbf{p}^{\text{u}}(t), p^{\text{H}}(t)] = [p_1^{\text{u}}(t), p_2^{\text{u}}(t), \cdots, p_K^{\text{u}}(t), p^{\text{H}}(t)]^T$ represent an instance of transmit power vector at time $t$. Our proposed power control scheme for controlling the downlink energy harvesting transmit power by the HBS is:

$$p^{\text{H}}(t+1) = f^{\text{H}}(\mathbf{p}(t)) = \min\{\overline{p}^{\text{H}}, \max_{i \in \mathcal{K}} \left( \frac{p_i^{\text{u}}(t)}{\epsilon \mu_i g_i} + p_i^{\min} \right)\}. \quad (15)$$

*2) Our proposed uplink power control schemes:* The power control scheme in (15) which is employed by the HBS for transferring energy to the UEs enables each UE to harvest its required energy for powering its circuit and transmitting its uplink information signal. Now, the question is how the UEs update their uplink transmit power levels so that Problem 1 or 2 is addressed. Inspired by two power control algorithms presented in [15] and [16] for traditional wireless networks without energy harvesting and IBFD capabilities, we propose two power control schemes to address Problems 1 and 2 in IBFD-EHWNs in a distributed manner, respectively, as explained in what follows.

To address Problem 1, we propose that each UE iteratively tracks its target-SINR by updating its uplink transmit power according to the following power updating scheme:

$$p_i^{\text{u}}(t+1) = f_i^{\text{u}}(\mathbf{p}(t)) = \min\{\overline{p}_i^{\text{u}}, \widehat{\gamma}_i \frac{\sum_{j \neq i} h_j p_j^{\text{u}}(t) + \delta p^{\text{H}}(t) + \sigma^2}{h_i}\}, \quad (16)$$

and the HBS updates its downlink energy harvesting transmit power according to the power update function (15). Let us call the transmit power update schemes proposed in (15) and (16) for controlling the transmit power levels of the HBS and the UEs, the **TPCEH algorithm**.

In Problem 2, for enhancing the aggregate throughput, it is required that the UEs with good channels transmit at high power levels and the UEs with poor channels transmit at low power levels [16]. By exploiting this fact, and inspired by the algorithm proposed in [16] for enhancing the aggregate throughput in conventional wireless networks, we propose the following power updating scheme for the UEs to address Problem 2,

$$p_i^{\text{u}}(t+1) = f_i^{\text{u}}(\mathbf{p}(t)) = \min\{\overline{p}_i^{\text{u}}, \eta_i \frac{h_i}{\sum_{j \neq i} h_j p_j^{\text{u}}(t) + \delta p^{\text{H}}(t) + \sigma^2}\}, \quad (17)$$

where $\eta_i$ is a non-negative constant for the $i$th UE, called the target signal-interference product [16]. The larger the value of $\eta_i$, the higher the transmitted power. Therefore, $\eta_i$ determines how the $i$th UE is eager to transmit. When all the UEs employ the power control scheme (17), the UEs that experience low (high) interference and have high (low) channel gain, transmit at higher (lower) power levels, which results in a significantly enhanced aggregate throughput [16]. Here, the interference includes interference from the other UEs and self-interference. We call the transmit power update schemes proposed in (15) and (17) for controlling the transmit power levels of the HBS and the UEs, the **OPCEH algorithm**.

*B. Convergence Analysis*

In this section, we show that our two proposed distributed power control schemes given by (15), (16) and (15), (17), respectively, converge to their unique fixed points. To prove the convergence, first we show that our proposed power update functions are two-sided scalable. Then we use it to prove that our proposed TPCEH and OPCEH schemes converge to their corresponding unique fixed points. To do this, we use two-sided scalable framework developed in [17] presented in the following definition.

**Definition 1.** An iterative function $\mathbf{f}(\mathbf{p})$ is said to be a two-sided scalable function if for all $a > 1$, $(1/a)\mathbf{p} \leq \mathbf{p}' \leq a\mathbf{p}$ implies $(1/a)\mathbf{f}(\mathbf{p}) \leq \mathbf{f}(\mathbf{p}') \leq a\mathbf{f}(\mathbf{p})$ [17].

**Lemma 2.** Let $\mathbf{f}(\mathbf{p}) = [\mathbf{f}^{\text{u}}(\mathbf{p}), f^{\text{H}}(\mathbf{p})]$ denote the transmit power update function, where $f^{\text{H}}(\mathbf{p})$ is given by (15) and $\mathbf{f}^{\text{u}}(\mathbf{p}) = [f_1^{\text{u}}(\mathbf{p}), f_2^{\text{u}}(\mathbf{p}), \cdots, f_K^{\text{u}}(\mathbf{p})]$ where $f_i^{\text{u}}(\mathbf{p})$ for all $i \in \mathcal{K}$ is given by (16) or (17). The transmit power update function $\mathbf{f}(\mathbf{p})$ is a two-sided scalable function.

*Proof.* First, we show that $f^{\text{H}}(\mathbf{p})$ proposed in (15) is a two-sided scalable function. Given $a > 1$, from $(1/a)\mathbf{p} = (1/a)[\mathbf{p}^{\text{u}}, p^{\text{H}}] \leq \mathbf{p}' = [\mathbf{p}'^{\text{u}}, p'^{\text{H}}] \leq a\mathbf{p} = a[\mathbf{p}^{\text{u}}, p^{\text{H}}]$, we have:

$$(1/a) \left( \max_{i \in \mathcal{K}} \left( \frac{p_i^{\text{u}}}{\epsilon \mu_i g_i} + p_i^{\min} \right) \right)$$
$$\leq \max_{i \in \mathcal{K}} \left( \frac{p_i'^{\text{u}}}{\epsilon \mu_i g_i} + p_i^{\min} \right)$$
$$\leq a \left( \max_{i \in \mathcal{K}} \left( \frac{p_i^{\text{u}}}{\epsilon \mu_i g_i} + p_i^{\min} \right) \right), \quad \forall a > 1. \quad (18)$$

This implies $(1/a)f^{\text{H}}(\mathbf{p}) \leq f^{\text{H}}(\mathbf{p}') \leq af^{\text{H}}(\mathbf{p})$. Now, we show that $\mathbf{f}^{\text{u}}(\mathbf{p})$ proposed in (16) is a two-sided scalable function. It has the following properties:

- Monotonicity: From $\mathbf{p} = [\mathbf{p}^{\text{u}}, p^{\text{H}}] \leq \mathbf{p}' = [\mathbf{p}'^{\text{u}}, p'^{\text{H}}]$, we have: $\widehat{\gamma}_i \frac{\sum_{j \neq i} h_j p_j^{\text{u}} + \delta p^{\text{H}} + \sigma^2}{h_i} \leq \widehat{\gamma}_i \frac{\sum_{j \neq i} h_j p_j'^{\text{u}} + \delta p'^{\text{H}} + \sigma^2}{h_i}$, which implies $\mathbf{f}^{\text{u}}(\mathbf{p}) \leq \mathbf{f}^{\text{u}}(\mathbf{p}')$.
- Scalability: For all $a > 1$, we have: $\widehat{\gamma}_i \frac{\sum_{j \neq i} ah_j p_j^{\text{u}} + a\delta p^{\text{H}} + \sigma^2}{h_i} < a\widehat{\gamma}_i \frac{\sum_{j \neq i} h_j p_j^{\text{u}} + \delta p^{\text{H}} + \sigma^2}{h_i}$, which implies $\mathbf{f}^{\text{u}}(a\mathbf{p}) < a\mathbf{f}^{\text{u}}(\mathbf{p})$.

It was shown in [17] that if $\mathbf{f}(\mathbf{p})$ has the above two properties, it is a two-sided scalable function. Finally, we show that $\mathbf{f}^{\text{u}}(\mathbf{p})$ proposed in (17) is a two-sided scalable function. It has the following properties:

- Type-II monotonicity: From $\mathbf{p} = [\mathbf{p}^{\text{u}}, p^{\text{H}}] \leq \mathbf{p}' = [\mathbf{p}'^{\text{u}}, p'^{\text{H}}]$, we have: $\eta_i \frac{h_i}{\sum_{j \neq i} h_j p_j^{\text{u}} + \delta p^{\text{H}} + \sigma^2} \geq \eta_i \frac{h_i}{\sum_{j \neq i} h_j p_j'^{\text{u}} + \delta p'^{\text{H}} + \sigma^2}$, which implies $\mathbf{f}^{\text{u}}(\mathbf{p}) \geq \mathbf{f}^{\text{u}}(\mathbf{p}')$.

- Type-II scalability: For all $a > 1$, we have: $\eta_i \frac{h_i}{\sum_{j \neq i} a h_j p_j^{\text{u}} + a\delta p^{\text{H}} + \sigma^2} > (1/a)\eta_i \frac{h_i}{\sum_{j \neq i} h_j p_j^{\text{u}} + \delta p^{\text{H}} + \sigma^2}$, which implies $\mathbf{f}^{\text{u}}(a\mathbf{p}) > (1/a)\mathbf{f}^{\text{u}}(\mathbf{p})$.

It was shown in [17] that if $\mathbf{f}(\mathbf{p})$ has the above two properties, it is a two-sided scalable function. Thus the transmit power update function $\mathbf{f}(\mathbf{p})$ is a two-sided scalable function. □

**Theorem 1.** The transmit power update function $\mathbf{f}(\mathbf{p}) = [\mathbf{f}^{\text{u}}(\mathbf{p}), f^{\text{H}}(\mathbf{p})]$, in which $f^{\text{H}}(\mathbf{p})$ is given by (15) and $\mathbf{f}^{\text{u}}(\mathbf{p}) = [f_1^{\text{u}}(\mathbf{p}), f_2^{\text{u}}(\mathbf{p}), \cdots, f_K^{\text{u}}(\mathbf{p})]$ with $f_i^{\text{u}}(\mathbf{p})$ given by (16) or (17) for all $i \in \mathcal{K}$, has a unique fixed point. Furthermore, given any initial power vector, our proposed power control scheme $\mathbf{p}(t+1) = \mathbf{f}(\mathbf{p}(t))$ converges to it.

*Proof.* It was shown in [17, Theorem 14] that given a two sided scalable function $\mathbf{f}(\mathbf{p})$, if there exists $\mathbf{l}, \mathbf{u} > 0$ such that $\mathbf{l} \leq \mathbf{f}(\mathbf{p}) \leq \mathbf{u}$ for all $\mathbf{p}$, then a unique fixed point exists and the power updating $\mathbf{p}(t+1) = \mathbf{f}(\mathbf{p}(t))$ converges to it. For our proposed transmit power update function $\mathbf{f}(\mathbf{p})$, since $0 < f_i^{\text{u}}(\mathbf{p}) \leq \overline{p}_i^{\text{u}} \ \forall i \in \mathcal{K}$ and $0 < f^{\text{H}}(\mathbf{p}) \leq \overline{p}^{\text{H}}$, then by considering $\mathbf{l} = [l_1, l_2, \cdots, l_K, l_{\text{HBS}}]$ where $l_i = q^{\min}, \forall i \in \{1, 2, \cdots, K, \text{HBS}\}$ in which $q^{\min} \to 0^+$ and $\mathbf{u} = [u_1, u_2, \cdots, u_K, u_{\text{HBS}}]$ where $u_i = \overline{p}_i^{\text{u}} \ \forall i \in \mathcal{K}$ and $u_{\text{HBS}} = \overline{p}^{\text{H}}$, we have $\mathbf{l} \leq \mathbf{f}(\mathbf{p}) \leq \mathbf{u}$ for all $\mathbf{p}$. Therefore, there exists a unique fixed point, to which our proposed power control scheme converges to. □

### C. Optimality Analysis

We now show that our proposed TPCEH algorithm converges to the optimal solution of Problem 1.

**Theorem 2.** Our proposed TPCEH (given by (15) and (16)) converges to the optimal solution of Problem 1 (i.e., its corresponding fixed point is the same as the optimal solution of Problem 1).

*Proof.* See the Appendix. □

In contrast to Problem 1, which is convex and optimally solved by our proposed TPCEH scheme, Problem 2 is non-convex due to non-convexity of its objective function, which makes it hard to optimally solve it in either a centralized or a distributed manner. However, in [34] and [35], some properties of the optimal solution for aggregate throughput maximization problem have been provided. Generally, it has been shown that in the optimal point, the UEs with better channels transmit at higher powers in comparison to the UEs with poor channels, as is the case in the OPC algorithm [16] for traditional cellular networks and our proposed OPCEH scheme for IBFD-EHWNs. Indeed, although our proposed distributed OPCEH scheme does not guarantee the optimal solution of Problem 2, it significantly improves the aggregate throughput to the same extent of improvement achieved by the OPC algorithm [16] in traditional cellular networks (without energy harvesting capability) as compared to the other power control algorithms such as the TPC algorithm [15]. This will be demonstrated via simulations in the next section.

## IV. NUMERICAL RESULTS

In this section, the performance of our proposed TPCEH and OPCEH schemes are evaluated and compared with other algorithms. We consider a single-cell system where UEs are generated uniformly within a square cell. For each UE, the channel gains are assumed to be the same for the uplink and the downlink, and are modeled as $g_i = h_i = k d_i^{-3}$, where $d_i$ is the distance between the $i$th UE and the HBS, $k$ is the attenuation factor that represents power variations set to $k = 0.09$. The upper bound on the transmit power for all UEs ($\overline{p}_i^{\text{u}}$) is 30 dBm and the amount of the energy harvesting transmit power by the HBS can be 40 dBm at most (i.e., $\overline{p}^{\text{H}} = 40$ dBm). Similar to [23], the parameters in the linear power model are set to $\epsilon = 0.2$, $p_{\text{HBS}}^{\text{Sta}} = 27$ dBm, $p_{\text{HBS}}^{\text{Dyn}} = 38$ dBm, $p_i^{\text{Sta}} = 20$ dBm, and $p_i^{\text{Dyn}} = 26$ dBm. Since the HBS and the UEs operate in IBFD mode by using two antennas, the number of antennas is 2 (i.e., $N_{\text{HBS}} = N_i = 2$). The AWGN power at the HBS receiver, i.e., $\sigma^2$, is assumed to be -113 dBm. In our simulations, the values of $\eta_i$ in OPCEH algorithm for all the UEs is assumed to be $\eta_i = 1$ (as in [17] and [18]). The energy harvesting efficiency ($\mu_i$) for each UE is assumed to be uniformly distributed within the interval $(0, 1)$.

We first consider a system with fixed UEs in the cell for a single snapshot to study the convergence behavior of our proposed schemes. Then, we consider the fixed-UEs system for different snapshot to investigate the effect of different parameters such as effective self-interference coefficient and cell side length on the proposed schemes and we investigate how they work, as compared to other algorithms (i.e., TPC and OPC). Finally, we consider a system with mobile UEs to investigate how our proposed schemes work when UEs are moving and may have not sufficient energy.

### A. Single Snapshot

To evaluate TPCEH and OPCEH algorithms, we consider five fixed UEs within a single cell with the side length of 50m where their distance vector is $\mathbf{d} = [41, 25, 37, 16, 8]^T$ m, in which each element is the distance of the corresponding UE from HBS. In our simulations, for TPCEH algorithm, we consider the target SINR vector of $\hat{\boldsymbol{\gamma}} = [0.04, 0.05, 0.07, 0.08, 0.1]$, in which each element is the target SINR of the corresponding UE. We assume that the effective self-interference coefficient (i.e., $\delta$) is -120 dB. Figs. 2 and 3 illustrate the total power consumption at UEs and HBS and the received SINR for the UEs in each iteration when our proposed TPCEH and OPCEH schemes are applied, respectively. In Fig. 2 we observe that despite the fact that the UEs have non-available permanent battery, they obtain their minimum target SINR (outage=0) by harvesting energy according to the TPCEH algorithm. In OPCEH (see Fig. 3), the transmit power of the UE with the best channel (UE 5) is high and it reaches a high SINR, and other UEs reach very low SINRs, which results in a significant improvement in the aggregate throughput [16], [17]. In both Figs. 2 and 3, the energy required by the UEs is harvested from the energy transmitted by the HBS well adjusted by using our proposed scheme (15). As seen in Figs. 2 and 3, the total power consumed by the UEs and the HBS converge to a fixed point.

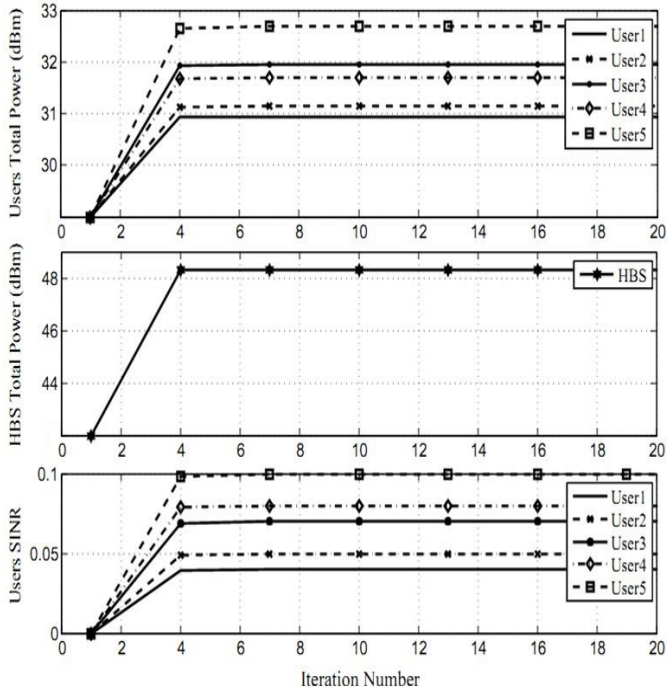

Fig. 2. Total power consumption ($\frac{1}{\epsilon}p_i^u + p_i^{cir}$) and SINRs for each UE and total power consumption at HBS ($\frac{1}{\epsilon}p^H + p_{HBS}^{cir}$) in TPCEH.

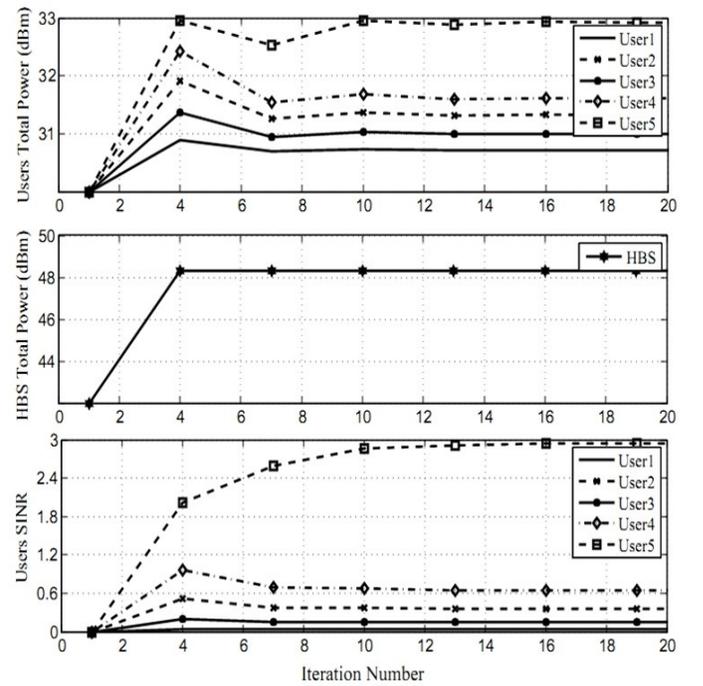

Fig. 3. Total power consumption ($\frac{1}{\epsilon}p_i^u + p_i^{cir}$) and SINRs for each UE and total power consumption at HBS ($\frac{1}{\epsilon}p^H + p_{HBS}^{cir}$) in OPCEH.

### B. Different Snapshots

Now, we first investigate the effect of different parameters such as effective self-interference coefficient and cell side length on the proposed schemes. Then, we compare our proposed TPCEH and OPCEH schemes with TPC [15] and OPC [16] algorithms, respectively. Since, to the best of our knowledge, there are no distributed power control schemes for IBFD-EHWN, we compare our proposed schemes only with the TPC and OPC algorithms which are already proposed for the traditional cellular networks without energy harvesting and IBFD capabilities. We show that our proposed schemes approach the performance of the TPC and OPC assuming IBFD capability and non-available permanent battery. In all scenarios, we consider 5000 independent snapshots with random locations for the UEs in each snapshot and the numerical results for each algorithm are obtained by averaging over all independent snapshots. In our simulations, for TPCEH algorithm, we consider target SINR of $\hat{\gamma}_i = 0.05$ for all UEs.

*1) Fixed UEs:* We first consider the system with fixed UEs in the cell. We investigate the effect of effective self-interference coefficient (i.e., $\delta$) on TPCEH and OPCEH for different cell side lengths. Fig. 4 illustrates the energy harvesting transmit power by the HBS and the average received SINR of the UEs versus the effective self-interference coefficient for different cell side lengths in TPCEH. For a fixed $\delta$, for example -80 dB, we see that increasing cell side length results in decreased SINR for the UEs. We observe that the UEs reach their minimum target SINRs with a lower value of $\delta$ as the cell side length increases. With an increasing cell side length, the effective self-interference coefficient should decrease to maintain the minimum target SINR of the UEs. For example, if the cell side length is 40m, the UEs can reach their minimum target SINRs when $\delta = -100$ dB. However, if the cell side length is 60m, the UEs can reach their minimum target SINRs with $\delta = -120$ dB. As seen in Fig. 4, increasing the effective self-interference coefficient causes the energy harvesting transmit power by the HBS to increase for different cell side lengths. For lower values of $\delta$, less energy harvesting transmit power is needed in small cell side lengths. We also observe that increasing $\delta$ causes the SINR of the UEs to decrease in TPCEH algorithm.

Fig. 5 illustrates the average system throughput versus the effective self-interference for different cell side lengths in OPCEH. It is observed that the system throughput decreases as the cell side length increases. From this figure, we also see that small values for the effective self-interference results in an increased system throughput in OPCEH. The results of Figs. 4 and 5 imply that the energy harvesting and IBFD capabilities are suitable for short distance communications such as networks with a small cell side length (i.e., femtocell networks) and D2D communications. In general, this observation (the suitability of IBFD capability for short distance communications) has been already made in [31] and [32] for different system models and resource allocation problems.

Now, we compare the proposed TPCEH and OPCEH schemes with the TPC and OPC algorithms, respectively.

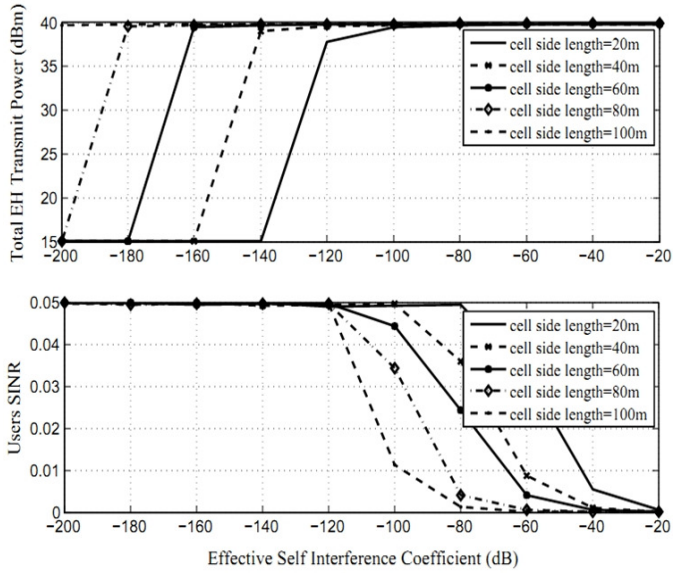

Fig. 4. Average energy harvesting transmit power and average UEs' SINR versus effective self-interference coefficient ($\delta$) for different cell side lengths in TPCEH.

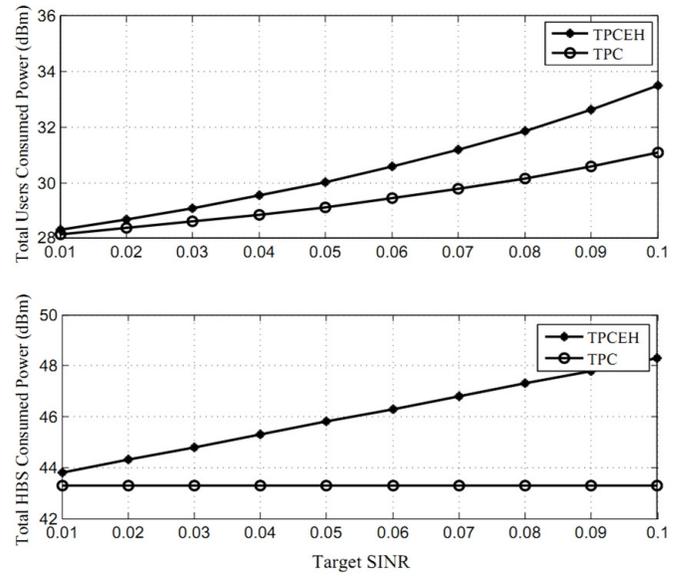

Fig. 6. Average total power consumed by UEs ($\sum_{i \in \mathcal{K}} \frac{1}{\epsilon} p_i^u + p_i^{cir}$) and average power consumed by HBS ($\frac{1}{\epsilon} p^H + p_{HBS}^{cir}$) versus target SINR in TPC and TPCEH.

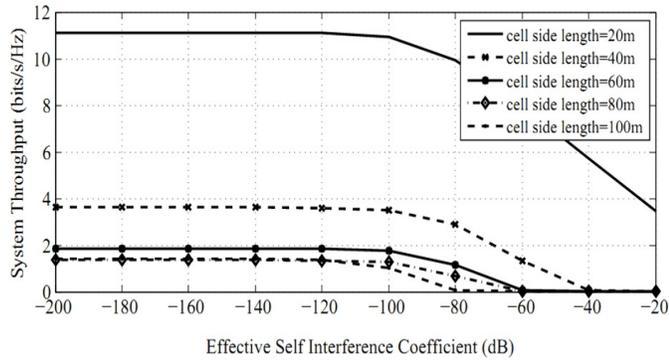

Fig. 5. Average system throughput versus effective self-interference coefficient ($\delta$) for different cell side lengths in OPCEH

The TPC and OPC algorithms operate in wireless cellular networks without energy harvesting capability. In fact, for TPC and OPC, it is assumed that the UEs are equipped with a permanent battery. While the UEs have no battery in our proposed TPCEH and OPCEH, and they provide their required energy by harvesting energy from the HBS. We assume that the cell side length is 50 m and the effective self-interference coefficient (i.e., $\delta$) is -120 dB. Our comparisons are made for different values of target-SINRs for the UEs and different total number of UEs.

Fig. 6 illustrates the effect of target SINR on the total consumed power by the UEs and the HBS under TPC and TPCEH algorithms. As expected, it is observed that the total power consumed by UEs for TPC and TPCEH increases when the target SINR increases. A higher target SINR requires higher transmission powers for the UEs. Therefore, they need more energy, which in turn increases the needed energy harvesting transmit power by the HBS leading to an increased total power consumed by the HBS in TPCEH. However, since the TPC algorithm [15] operates in a system without energy harvesting capability, the HBS does not transmit energy signals and the total power consumed by the HBS for TPC (which only includes the HBS circuit power) remains fixed. From Fig. 6, we also see that the total power consumed by the UEs in TPCEH is more than that of TPC. The reason is that self-interference (caused by IBFD capability) enables the UEs to transmit at higher power to obtain the minimum target SINR in TPCEH, as opposed to the TPC in which the UEs operate in half-duplex mode.

Fig. 7 shows the effect of the number of UEs on the throughput of the system and the total power consumed by UEs and HBS under OPC and OPCEH algorithms. It is observed that increased number of UEs ($K$) results in a decreased system throughput. A higher number of UEs indicates that the interference increases, and thus throughput of the system decreases. It is also observed that the total power consumed by the UEs increases when the number of UEs increases. The important observation is that the throughput and also the total power consumed by the UEs for both OPC and OPCEH are the same. In fact, by applying our proposed OPCEH scheme in energy harvesting wireless networks with IBFD capability and non-available permanent battery for the UEs, the system throughput is equal to that of the OPC algorithm in traditional wireless networks without IBFD and energy harvesting capabilities in which the UEs have permanent batteries. The reason is that the UE with the best channel transmits at a high power and obtains a high throughput in both algorithms. As we see in Fig. 7, the total power consumed by the HBS increases when the number of UEs increases in OPCEH due to an increasing energy harvesting transmit power at the HBS. However, since the OPC algorithm [16] operates in a system without energy harvesting capability, the HBS does not transmit energy signals and the total power consumed by the HBS for OPC (which only includes the HBS circuit power) remains fixed.

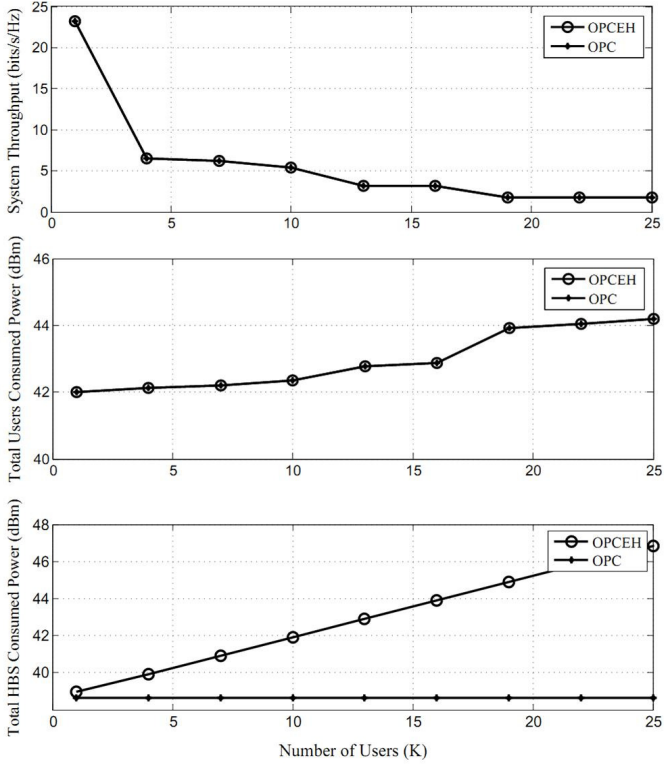

Fig. 7. Average system throughput and average total UEs transmit power ($\sum_{i\in\mathcal{K}} p_i^{\text{u}}$) and average energy harvesting transmit power ($p^{\text{H}}$) versus the number UEs in OPC and OPCEH.

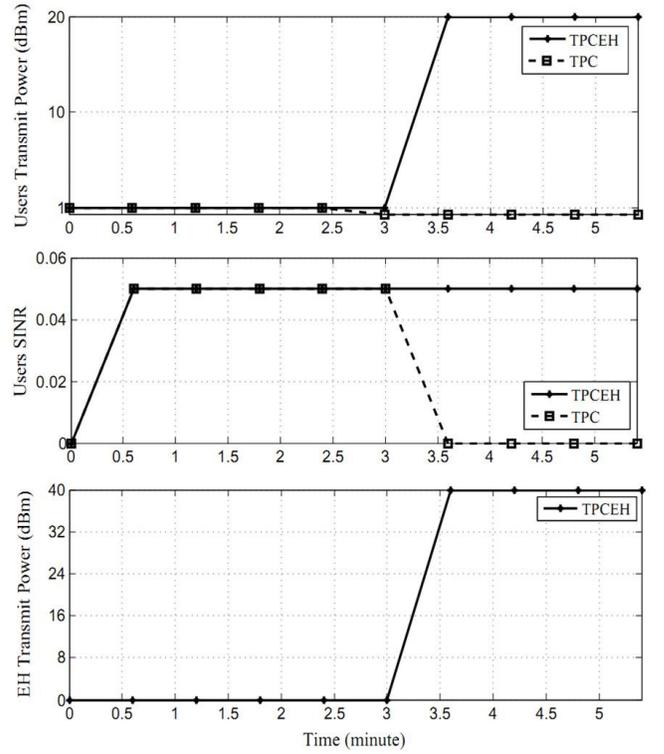

Fig. 8. Average UEs and energy harvesting transmit power and average UEs SINR versus time in TPC and TPCEH with moving UEs.

*2) Mobile UEs:* At last, to show how TPCEH works when the UEs do not have sufficient energy to transmit their data, we assume they have a battery with initial energy equal to 1 $\mu$J. Assuming that the UEs move in the cell, suppose that at $t = 0$, the UEs start moving from the beginning of the cell toward its end and come back along a straight line at a uniform speed of 5 km/h. Each UE updates its transmit power every 1 ms. We consider the target SINR of $\hat{\gamma}_i = 0.05$ for all the UEs. As we observe in Fig. 8, when the UEs employ TPC, their initial energy runs out after three minutes, after which, they cannot transmit their data and their SINR become zero. However, when the UEs employ TPCEH, they harvest energy from the HBS after exhausting their initial energy at $t = 3$ s, so they can continue to transmit data and obtain the minimum target SINR. The HBS starts to transmit energy to the UEs after their initial energy runs out at $t = 3$ s. A identical observation is made for the OPCEH scheme which is not included due to the similarity in the simulation results.

## V. Conclusions and Future Work

We have studied the problems of minimizing the aggregate power subject to the UEs' target-SINRs and the energy harvesting constraints, and maximizing the aggregate throughput subject to the energy harvesting constraint, both in IBFD-EHWNs. We addressed these problems by proposing two distributed power control schemes for controlling the uplink transmit power by the UEs and the downlink energy harvesting signal power by the HBS. We proved that both TPCEH and OPCEH schemes converge to their corresponding unique fixed point starting from any arbitrary initial transmit power. Furthermore, We have also shown that our proposed TPCEH algorithm guarantees that each UE harvests its required energy to reach its target SINR, while the minimum aggregate power is consumed by the HBS and the UEs. In this paper, we assumed that the energy source is dedicated and the HBS transfers energy to the UEs. Considering the leakage energy (i.e., inter-user interference) between different UEs in the uplink as the energy source in addition to the transferred energy by the HBS is a complicating scenario to be considered in our future works.

## Appendix
### Proof of the Theorem 2

To prove the theorem 2, we employ the Fast-Lipschitz Optimization (FLO) method developed in [20], which is presented in the following statements.

**Definition 2.** A problem is said to have Fast-Lipschitz form if it can be written as

$$\underset{\mathbf{x}}{\text{maximize}} \quad f_0(\mathbf{x}) \quad (19)$$

$$\text{subject to} \quad x_i \leq f_i(\mathbf{x}), \quad \forall i = 1, \cdots, l \quad (19\text{-}1)$$

$$\quad x_i \leq h_i(\mathbf{x}), \quad \forall i = l+1, \cdots, n \quad (19\text{-}2)$$

$$\mathbf{x} \in \mathcal{D}, \quad (19\text{-}3)$$

where $\mathcal{D} \subset \mathcal{R}^n$ is the bounded space of $[x_1^{min}, x_1^{max}] \times [x_2^{min}, x_2^{max}] \cdots [x_n^{min}, x_n^{max}]$ [20].

**Definition 3.** A problem is said to be Fast-Lipschitz if it can be written in Fast-Lipschitz form while satisfying the following properties:

$$\nabla f_0(\mathbf{x}) > 0,$$
$$\nabla \mathbf{F}(\mathbf{x}) \geq 0,$$
$$\|\nabla \mathbf{F}(\mathbf{x})\|_\infty < 1, \quad (20)$$

where $\mathbf{F}(\mathbf{x}) = [f_1(\mathbf{x}) \cdots f_l(\mathbf{x}) h_{l+1}(\mathbf{x}) \cdots h_n(\mathbf{x})]^T$. Here $\nabla$ denotes the gradient operator and $\nabla f(x)$ is the transpose of the Jacobian matrix, i.e., $[\nabla f(x)]_{ij} = \dfrac{\partial f_j(x)}{\partial x_i}$, whereas $\nabla_i f(x)$ denotes the $i$th row of $\nabla f(x)$. For a matrix $\mathbf{A} \in \mathcal{R}^{n \times n}$ we use the induced norm defined as $\|\mathbf{A}\|_\infty = \max_{j=1}^n \sum_{i=1}^n |a_{ij}|$ [19].

**Proposition 1.** The optimal solution of a Fast-Lipschitz problem is obtained in a distributed manner by iterating the constraint functions as follows

$$x_i(t+1) = [f_i(x_i(t))]^{\mathcal{D}}, \quad \forall i = 1, \cdots, l$$
$$x_i(t+1) = [h_i(x_i(t))]^{\mathcal{D}}, \quad \forall i = l+1, \cdots, n, \quad (21)$$

where $[x_i]^{\mathcal{D}} = x_i$ if $x_i \in [x_i^{min}, x_i^{max}]$, $[x_i]^{\mathcal{D}} = x_i^{min}$ if $x_i < x_i^{min}$, or $[x_i]^{\mathcal{D}} = x_i^{max}$ if $x_i > x_i^{max}$ [20].

First, we write Problem 1 in Fast-Lipschitz form. By replacing $\gamma_i(\mathbf{p})$ from (9) in the constraint $\gamma_i(\mathbf{p}) \geq \widehat{\gamma}_i, \forall i \in \mathcal{K}$ of Problem 1, we have:

$$p_i^{\mathrm{u}} \geq \widehat{\gamma}_i \frac{\sum_{j \in \mathcal{K}, j \neq i} h_j p_j^{\mathrm{u}} + \delta p^{\mathrm{H}} + \sigma^2}{h_i}, \quad \forall i \in \mathcal{K}. \quad (22)$$

By adding $\widehat{\gamma}_i p_i^{\mathrm{u}}$ to both sides of Equation (22), we have

$$p_i^{\mathrm{u}} \geq \frac{\widehat{\gamma}_i}{(1+\widehat{\gamma}_i)h_i} \left( \sum_{l \in \mathcal{K}} h_l p_l^{\mathrm{u}} + \delta p^{\mathrm{H}} + \sigma^2 \right), \quad \forall i \in \mathcal{K}. \quad (23)$$

On the other hand, the constraint $p^{\mathrm{H}} \geq \dfrac{p_i^{\mathrm{u}}}{\epsilon \mu_i g_i} + p_i^{\min}, \forall i \in \mathcal{K}$ in Problem 1 should hold. So, we have

$$p^{\mathrm{H}} \geq \max_{i \in \mathcal{K}} \left( \frac{p_i^{\mathrm{u}}}{\epsilon \mu_i g_i} + p_i^{\min} \right). \quad (24)$$

Therefore, if $\gamma_i(\mathbf{p}) \geq \widehat{\gamma}_i, \forall i \in \mathcal{K}$, from (23) and (24) we conclude:

$$p^{\mathrm{H}} \geq \max_{i \in \mathcal{K}} \left( \frac{\widehat{\gamma}_i}{(1+\widehat{\gamma}_i)\epsilon h_i g_i \mu_i} \left( \sum_{l \in \mathcal{K}} h_l p_l^{\mathrm{u}} + \delta p^{\mathrm{H}} + \sigma^2 \right) + p_i^{\min} \right). \quad (25)$$

Let us define $\alpha_i = \dfrac{\widehat{\gamma}_i}{(1+\widehat{\gamma}_i)\epsilon h_i g_i \mu_i}$. So, we can rewrite (25) as

$$p^{\mathrm{H}} \geq \max_{i \in \mathcal{K}} \left( \alpha_i \left( \sum_{l \in \mathcal{K}} h_l p_l^{\mathrm{u}} + \delta p^{\mathrm{H}} + \sigma^2 \right) + p_i^{\min} \right). \quad (26)$$

Motivated by example 2 in [19], the change of variable $\mathbf{x} = -\mathbf{p}$ transforms Problem 1 to Fast-Lipschitz form as follows

$$\begin{aligned}\underset{\mathbf{x}}{\text{maximize}} \quad & f_0(\mathbf{x}) = -g_0(-\mathbf{x}) & (27)\\ \text{subject to} \quad & x_i^{\mathrm{u}} \leq f_i(\mathbf{x}) = -g_i(-\mathbf{x}), \quad \forall i \in \mathcal{K} & (27\text{-}1)\\ & x^{\mathrm{H}} \leq h(\mathbf{x}) = -z(-\mathbf{x}), & (27\text{-}2)\\ & \mathbf{x} \in \mathcal{D}, & (27\text{-}3)\end{aligned}$$

where

$$g_0(\mathbf{x}) = \frac{1}{\epsilon}\left(\sum_{i=1}^K x_i^{\mathrm{u}} + x^{\mathrm{H}}\right) + \sum_{i=1}^K p_i^{\mathrm{cir}} + p_{\mathrm{HBS}}^{\mathrm{cir}}$$

$$g_i(\mathbf{x}) = \frac{\widehat{\gamma}_i}{(1+\widehat{\gamma}_i)h_i}\left(\sum_{l\in\mathcal{K}} h_l x_l^{\mathrm{u}} + \delta x^{\mathrm{H}} + \sigma^2\right),$$

$$z(\mathbf{x}) = \max_{i\in\mathcal{K}}\left(\alpha_i\left(\sum_{l\in\mathcal{K}} h_l x_l^{\mathrm{u}} + \delta x^{\mathrm{H}} + \sigma^2\right) + p_i^{\min}\right),$$

and $\mathcal{D} \subset \mathcal{R}^n$ is the bounded space of $[0, \overline{x}_1^{\mathrm{u}}] \times [0, \overline{x}_2^{\mathrm{u}}] \cdots [0, \overline{x}^{\mathrm{H}}]$.

The problem stated in (27) is the Fast Lipschitz form of Problem 1. Since the right side of Equation (26) is a piecewise-linear differentiable function [36], and as it can be easily seen that $\nabla g_0(\mathbf{x}) > 0$, $\nabla \mathbf{G}(\mathbf{x}) \geq 0$ and $\|\nabla \mathbf{G}(\mathbf{x})\|_\infty < 1$, where $\mathbf{G}(\mathbf{x}) = [g_1(\mathbf{x}), \cdots g_K(\mathbf{x}), z(\mathbf{x})]^T$, we conclude that the problem given by (27) satisfies the properties stated in (20). Therefore, based on Proposition 1, the optimal solution is obtained in a distributed manner, by the following iterations:

$$x^{\mathrm{H}}(t+1) = \min\{\overline{p}^{\mathrm{H}}, \max_{i\in\mathcal{K}}(\alpha_i(\sum_{l\in\mathcal{K}} h_l x_l^{\mathrm{u}} + \delta x^{\mathrm{H}} + \sigma^2) + p_i^{\min})\} \quad (28)$$

$$x_i^{\mathrm{u}}(t+1) = \min\{\overline{p}_i^{\mathrm{u}}, \frac{\widehat{\gamma}_i}{(1+\widehat{\gamma}_i)h_i}(\sum_{l\in\mathcal{K}} h_l x_l^{\mathrm{u}}(t) + \delta x^{\mathrm{H}}(t) + \sigma^2)\}, \forall i \in \mathcal{K}. \quad (29)$$

By implementing Equations (23) to (26) in a reverse order, we can show that Equations (28) and (29) are equivalent to Equations (15) and (16), respectively. This completes the proof. ∎

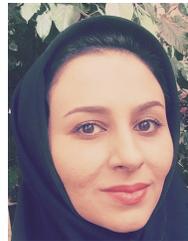

**Rojin Aslani** received her B.Sc. degree in Information Technology from University of Tabriz, Tabriz, Iran, in 2011 and her M.Sc. degree in Information Technology (Computer Networks) from Amirkabir University of Technology (AUT), Tehran, Iran, in 2013. She is currently pursuing the Ph.D. degree in Computer Engineering in Amirkabir University of Technology. Her current research area includes resource allocation in wireless networks.

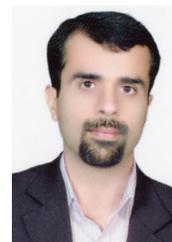

**Mehdi Rasti** (S'08-M'11) received his B.Sc. degree from Shiraz University, Shiraz, Iran, and the M.Sc. and Ph.D. degrees both from Tarbiat Modares University, Tehran, Iran, all in Electrical Engineering in 2001, 2003 and 2009, respectively. From November 2007 to November 2008, he was a visiting researcher at the Wireless@KTH, Royal Institute of Technology, Stockholm, Sweden. From September 2010 to July 2012 he was with Shiraz University of Technology, Shiraz, Iran, after that he joined the Department of Computer Engineering and Information Technology, Amirkabir University of Technology, Tehran, Iran, where he is now an assistant professor. From June 2013 to August 2013, and from July 2014 to August 2014 he was a visiting researcher in the Department of Electrical and Computer Engineering, University of Manitoba, Winnipeg, MB, Canada. His current research interests include radio resource allocation in wireless networks and network security.